\documentclass[12pt]{article}
\usepackage{amsfonts,color}
\usepackage{graphicx,verbatim}
\usepackage{amsmath,amssymb,amsthm}
\usepackage{amssymb}
\usepackage{mathtools}
\usepackage{mathrsfs}
\usepackage[english]{babel}
\usepackage[latin1]{inputenc}
\definecolor{DarkBlue}{rgb}{0.1,0,0.55}
\definecolor{DarkRed}{rgb}{0.8,0.1,0.1}
\def\e{\mathrm e}
\def\ri{\mathrm i}
\usepackage[height=10in, width=7in]{geometry}

\numberwithin{equation}{section}
\newcommand{\be}{\begin{equation}}
\newcommand{\ee}{\end{equation}}
\newcommand{\bea}{\begin{eqnarray}}
\newcommand{\eea}{\end{eqnarray}}
\newcommand{\ba}{\begin{array}}
\newcommand{\ea}{\end{array}}
\newcommand{\rf}[1] {(\ref{#1})}
\newcommand{\half}{\mbox{$\frac{1}{2}$}}
\newcommand{\eps}{\epsilon}

\def\ri{\mathrm{i}}

\def\e{\mathrm{e}}

\usepackage{times}

\title{Higher-Order Extensions of Weakly Viscous Dysthe Theory and a Phase-Lag Model for Nonlinear Mean-Flow Damping}

\author{C.M. Schober and A. Islas\\
University of Central Florida}
\date{}
\begin{document}

  \maketitle
  \begin{abstract}

    We extend the Carter-Govan multiple-scales analysis of weakly viscous, narrowband deep-water wave packets beyond Dysthe order within the potential-flow reduction of Dias, Dyachenko, and Zakharov (DDZ). We seek to determine whether this framework generates the complex multiplier $(1 + i\beta)$ used phenomenologically to modify the  nonlocal Dysthe mean-flow interaction.  Although order counting places a direct viscous carrier-mean interaction at sixth order, it does not exclude an indirect fifth-order contribution arising from viscosity dependent
    lower-order harmonics and nonlinear interactions. We therefore derive the first correction to the induced mean flow and the complete fifth-order first-harmonic solvability condition. The resulting nonlocal terms are derivative--dependent and contain no explicit viscosity, excluding the proposed indirect mechanism within the DDZ framework.
    At sixth order, a restricted calculation of the nonlinear viscous block isolates a direct carrier-mean contribution with the same operator structure as the imaginary component of the prescribed mean-flow correction. 
    Independently, a finite-adjustment-time model yields an exact,
    frequency dependent mean-flow response. Its low-frequency expansion produces the multiplier $ 1 + i\beta_{\mathrm{eff}}(\Omega)$, with
    $\beta_{\mathrm{eff}}(\Omega) =\Omega\tau.$ When $\Omega\tau = \mathcal O(\epsilon)$,
    the
    resulting phase-lag correction enters at fifth order, one order beyond the leading Dysthe mean-flow interaction.
  \end{abstract}
  \section{Introduction}

  The modulation of weakly nonlinear, narrowband deep-water surface-gravity
wave packets has been studied asymptotically since the late 1960s. Starting
from the inviscid, irrotational free-surface Euler equations, multiple-scales
expansions produce envelope models for the slow evolution of a nearly
monochromatic carrier. At the leading nonlinear modulation order, the cubic
nonlinear Schr\"odinger (NLS) equation governs the envelope
\cite{Z1968, HO1972} and describes the Benjamin-Feir sideband
instability of a uniform Stokes wavetrain \cite{BF1967}. At the next
order in wave steepness, Dysthe's fourth-order envelope equation adds
higher-order dispersive and nonlinear corrections and includes the nonlocal
feedback of the induced mean flow \cite{D1979}. These equations form the
asymptotic foundation for the weakly viscous extensions considered here.

  Within this hierarchy, the Dysthe equation provides a  more
realistic description of modulated deep-water wave groups and  gives improved agreement with laboratory observations. 
Comparisons 
with measurements of modulated periodic
Stokes wavetrains indicate that Dysthe simulations  captures essential macroscopic
features of the long-time dynamics of modulational instability (MI), including
complex and chaotic wave-group evolution.
However, damping must be incorporated separately to account for dissipative effects observed in laboratory experiments and field measurements
\cite{LM1985,DTKS2003,AHHS2001}.

Permanent frequency downshift is an important  irreversible features observed during the long-time evolution of modulationally unstable deep-water wave trains. It is characterized by a sustained shift in  spectral dominance from the carrier toward lower frequencies, rather than by a temporary redistribution between the sidebands \cite{LYRF1977}.  The growth of modulational instability is also accompanied by nonlinear focusing and the formation of large, steep wave groups, including rogue waves, making accurate prediction of both wave-group evolution and spectral change important for ship safety and offshore engineering
\cite{DKM2008}. The Dysthe equation captures asymmetric sideband growth and transient spectal shifts,  but its  approximate recurrence
prevents the shift from becoming permanent \cite{J1983,LM1985}.

This limitation motivated several dissipative extensions of Dysthe-type
equations. Of particular interest  here is the
Dysthe model with nonlinear mean-flow damping, in which the real
coefficient of the nonlocal Hilbert-transform mean-flow term is
replaced by the  complex factor $(1 + i\beta)$, with
$0 < \beta \ll 1$ 
 \cite{KO1995,IS2011}. In the nondimensional temporal-evolution form used in our earlier
studies, the model is
\begin{equation}
\ri u_t + u_{xx} + 2u|u|^2  + \epsilon\left[ 2u (1 + \ri\beta)\mathscr{H} \left(|u|^2_x\right) - 8\ri|u|^2 u_x + \frac{\ri}{2} u_{xxx} \right] = 0.
\label{1.1}
\end{equation}
Here $\epsilon \ll 1$ is a nondimensional steepness  parameter, $\beta$  
measures the strength of the nonlinear mean-flow damping, and
 $\mathscr{H}$ denotes the Hilbert transform for the 
periodic  envelope $u(x,t)$.
 Setting $\beta = 0$ recovers the Hamiltonian  Dysthe
 equation of Gramstad and Trulsen \cite{GT2011}.  The  imaginary
 component $i \beta$ represents dissipation acting specifically through the induced mean-flow interaction.  Although the Hilbert-transform
operator is nonlocal, the damping becomes most pronounced during  episodes of strong modulation, when the envelope intensity develops large spatial gradients. 

Although introduced as a phenomenological
correction rather than derived from
the governing water-wave equations, the mean-flow-damped Dysthe  model
reproduces
permanent frequency
downshift and has performed well in comparisons of reduced downshift models
\cite{IS2011,CHB2019}.
More recent work shows  that this form of damping suppresses disordered multimode evolution and promotes persistent, spectrally coherent soliton-like states from which sharply localized rogue waves emerge. Rogue-wave events in this model
are also closely associated with the onset of permanent downshift \cite{SI2025}.
These results make the asymptotic origin of
the prescribed complex mean-flow correction an important unresolved question.

Earlier higher-order envelope theories also considered dissipation acting through the carrier-induced mean flow. Fabrikant formulated coupled equations for a carrier envelope
  and an induced
  low-frequency  mode with distinct dissipation and showed that, under a quasistatic reduction, damping of the latter generates a nonconservative nonlinear derivative term in the carrier equation \cite{F1984}.
  Uchiyama and Kawahara incorporated an analogous contribution through a complex coefficient in a higher-order NLS model and found that its dissipative component produces an irreversible spectral
  shift \cite{UK1994}.
  Their proposed water-wave mechanism was heuristic, based on bottom friction acting on the induced long-wave component, and a systematic derivation from the governing equations was left open. Moreover, their correction is also local rather than the nonlocal
  Hilbert-transform interaction of deep-water Dysthe theory. These studies support the idea that dissipation can act through the carrier-induced mean flow, but they do not derive the specific complex correction considered here.

  A complementary approach is provided by the viscous Dysthe equation derived by Carter and Govan from the weak-viscosity formulation of Dias, Dyachenko, and Zakharov (DDZ) \cite{DDZ2008}. Its dissipative terms arise systematically from modified free-surface boundary conditions, and the model produces frequency downshift without explicitly invoking wind forcing or wave breaking. Numerical studies
  show that the spectral mean generally decreases and that the spectral peak
  can also shift downward  for suitable initial conditions, while  comparisons with laboratory experiments found good agreement in cases both with and without observed downshift \cite{CG2016}.
Subsequent work identified  viscous Dysthe and the nonlinear mean-flow damped Dysthe  as  among the most successful reduced descriptions of the measured spectral mean and peak \cite{CHB2019}. Weakly dissipative envelope models have also
improved predictions of long-range swell decay relative to nondissipative models, although nonlinear corrections were not clearly superior to their linearized counterparts in that setting
 \cite{ZC2021}. These results make the DDZ/Carter-Govan hierarchy a natural framework in which to investigate whether the prescribed nonlinear mean-flow correction has a systematic weak-viscosity origin.

 We therefore ask whether the DDZ/Carter-Govan hierarchy  beyond Dysthe order generates the complex correction to the induced  mean-flow interaction or produces
 additional nonlocal mean-flow terms. At Dysthe order, the coefficient of this interaction is real, leaving its higher-order viscous modification unresolved.

  Under the weak-viscosity scaling
$\bar\nu = \epsilon^2 \nu$,
order counting places the first direct viscous correction to the fourth-order Dysthe mean-flow interaction at sixth order. This observation does not resolve whether the same operator structure can arise indirectly at fifth order. Viscosity is already present in the lower-order envelope dynamics and harmonic corrections, and these quantities enter the nonlinear interactions contributing to the fifth-order first-harmonic solvability condition. An indirect fifth-order contribution with the same operator structure as the prescribed correction therefore remains possible until the
complete fifth-order calculation is carried out. 

The perturbation analysis below is carried out in the dimensional slow variables of the DDZ/Carter-Govan hierarchy.
The nondimensional mean-flow-damped damped Dysthe equation \rf{1.1} is included to identify the target complex correction; the comparison concerns its asymptotic order and operator structure rather than a term-by-term correspondence between the two formulations.

We first recover the DDZ/Carter-Govan perturbation hierarchy through Dysthe order and determine the additional harmonic corrections required at fifth-order. We then derive the first correction to the induced mean flow and the complete fifth-order first-harmonic solvability condition.  At  sixth order, we
isolate the direct carrier-mean contribution  from the nonlinear viscous block.

The fifth-order calculation yields new derivative-dependent nonlocal corrections but no viscosity-dependent modification of the leading Dysthe mean-flow operator. By contrast, the restricted sixth-order calculation identifies a direct contribution with the operator structure of the imaginary component of the prescribed correction, although its net coefficient in the complete sixth-order equation remains undetermined.

Independently of the DDZ asymptotic reduction, we introduce a finite-adjustment-time model for the induced mean flow. Its  frequency response contains an out-of-phase component that reduces at low frequency to the prescribed complex multiplier. Under the scaling $\Omega\tau = \mathcal O(\epsilon)$, the corresponding phase-lag correction enters at fifth order.

The remainder of the paper is organized as follows. Sections 2 and 3 review the governing equations and multiple-scales framework. Sections 4 to 6 develop the perturbation hierarchy through the fifth-order solvability condition, isolate the direct  sixth-order carrier-mean contirbution, and discuss the scope of the DDZ reduction. Section 7 analyzes the finite-adjustment-time model, and Section 8 summarizes the main conclusions.

\section{Background and Governing Equations}
We consider two-dimensional free-surface gravity waves in an incompressible, weakly viscous fluid of infinite depth. Let $x$ denote the horizontal coordinate, $z$ the vertical coordinate directed upward, $z=\eta(x,t)$ the free surface, and 
$\mathbf v = (u(x,z,t),w(x,z,t))$ the fluid velocity. The underlying viscous
free boundary problem is governed by the incompressible Navier-Stokes  equations \cite{J1997},

\bea
\mbox{Incompressibility} && \nabla\cdot \mathbf v = 0,\qquad -\infty<z<\eta(x,t), \nonumber\\ 
\mbox{Conservation of momentum} && \mathbf v_t + \left(\mathbf v\cdot\nabla\right)\mathbf v = -\frac{1}{\rho}\nabla p + \bar\nu\Delta \mathbf v + \mathbf g,
\qquad -\infty<z<\eta(x,t), \label{2.1}\\
 \mbox{Kinematic BC} && \eta_t + u \eta_x = w \qquad \mbox{at $z = \eta$},\nonumber\\
 \mbox{Dynamic BC} && -\left(p - p_0\right)\mathbf n + \underline{\underline{\mathbf\tau}}\cdot\mathbf n = 0 \quad\mbox{at $z = \eta$},\nonumber\\
 \mbox{Deep-water decay condition} &&|\mathbf v| \rightarrow 0 \qquad \mbox{as $z\rightarrow -\infty$}.\nonumber
\eea
Here $p(x,z,t)$ is the pressure, $\rho$ is the fluid density, $\bar\nu$ is the kinematic viscosity, $\mathbf g = (0,-g)$, and $\nabla=(\partial_x,\partial_z)$.
The vector $\mathbf n$ is the normal to the free surface and $\underline{\underline{\mathbf\tau}}$ is the viscous part of the stress tensor. 
The full viscous flow is not assumed to be exactly irrotational. In the inviscid formulation introduced below, the velocity is represented by a potential. In the Dias, Dyachenko, and Zakharov (DDZ) weak-viscosity reduction, the bulk flow remains potential, while the leading effect of a small rotational component near the free surface is incorporated through modified boundary conditions \cite{DDZ2008}.

\subsection{Inviscid Potential-Flow Formulation}
In the inviscid limit, we assume that the flow is irrotational. The velocity field can then be represented by a potential
 \(\phi(x,z,t)\), so that 
\be
\mathbf{v}  = \nabla\phi.
\ee
Neglecting viscosity  and integrating the momentum equation, the governing equations can be written as
\bea
\mbox{Laplace's equation} && \phi_{xx}+\phi_{zz}=0,\qquad -\infty<z<\eta(x,t),\nonumber\\
\mbox{Bernoulli equation} && \phi_t+\frac12|\nabla\phi|^2+g\eta=0,\qquad z=\eta(x,t),\nonumber\\
\mbox{Kinematic boundary condition} && \eta_t+\eta_x\phi_x=\phi_z,\qquad z=\eta(x,t),\nonumber\\
\mbox{Decay with depth} && |\nabla\phi|\rightarrow0,\qquad z\rightarrow-\infty.\nonumber 
\eea
These equations constitute the classical potential-flow formulation of the free-surface Euler equations and provide the inviscid basis for the weakly viscous reduction introduced next \cite{L1945}.

\subsection{Weakly Viscous Free-Surface Equations}

Dias, Dyachenko, and Zakharov showed that weak viscous effects can be incorporated through modified free surface boundary conditions while retaining a potential-flow description in the fluid interior \cite{DDZ2008}. Using a Helmholtz decomposition, they write the full velocity as the sum of  potential and rotational components,
\be
\mathbf v = \nabla \phi + \nabla\times \mathbf A.
\ee
For two-dimensional flow in the $x$-$z$ plane, the vector potential may be taken to have only a transverse component,  $\mathbf A = (0, A(x,z,t),0)$, so that the rotational velocity in the $x$-$z$ plane is $(-A_z,A_x)$. Equivalently
$\mathbf v = \nabla \phi + (-A_z,A_x)$.
The potential component is harmonic in the fluid interior, whereas the rotational component accounts for the vorticity required by the viscous free-surface conditions.

For a carrier mode with wavenumber $k$ and frequency $\omega$, the DDZ reduction
assumes $\frac{\bar\nu k^2}{|\omega|} << 1$.
This condition requires  the viscous damping rate to be small relative to the carrier frequency. Equivalently, the rotational motion is confined to a surface layer that is thin compared with the carrier wavelength. The rotational component is therefore asymptotically small  and may be eliminated to leading order, while its influence is retained in the free-surface boundary conditions.

Applying the Helmholtz decomposition to the linearized Navier-Stokes equations, DDZ obtain
\bea
A_x &=& 2\bar \nu \eta_{xx}\nonumber\\
p &=& p_0 + 2\rho\bar\nu\left(\phi_{zz} + A_{xz}\right).\nonumber
\eea

Because the rotational correction is already first order in $\bar\nu$,
its contribution through $\bar\nu A_{xz}$ is of higher order and is neglected
in the leading weak-viscosity approximation. The retained terms produce the viscous corrections $-2\bar{\nu}\phi_{zz}$ and $2\bar{\nu}\eta_{xx}$ in  the dynamic and kinematic boundary conditions, respectively.

DDZ then propose that these leading corrections retain the same differential form on a weakly nonlinear, small-slope free surface. The resulting quasi-potential system is
\bea
\phi_{xx}+\phi_{zz}
&=&
0,
\qquad
-\infty<z<\eta(x,t),
\label{2.4}
\\
\phi_t
+\frac{1}{2}|\nabla\phi|^2
+g\eta
&=&
-2\bar{\nu}\phi_{zz},
\qquad
z=\eta(x,t),
\label{2.5}
\\
\eta_t+\eta_x\phi_x
&=&
\phi_z+2\bar{\nu}\eta_{xx},
\qquad
z=\eta(x,t).
\label{2.6}\\
|\nabla\phi| & \rightarrow & 0,\qquad z\rightarrow-\infty.\label{2.7}
\eea

This is the weakly viscous system employed by Carter and Govan in their derivation of the viscous Dysthe equation \cite{CG2016}. The DDZ system is a reduced weak-viscosity model rather than the complete Navier-Stokes free-boundary problem. The rotational surface-layer component is eliminated asymptotically, and only its leading influence is retained through the two modified free-surface boundary conditions.

\subsection{Derivative Bernoulli Equation}

For the multiple-scales analysis, it is convenient to eliminate the explicit
occurence of the free-surface elevation $\eta$ from the dynamic boundary condition. We differentiate Bernoulli's equation along the moving free surface and
use  the kinematic boundary condition, together with its spatial derivatives, to eliminate derivatives of \(\eta\). This yields the derivative Bernoulli equation  \cite{D1979}
\be
\begin{split}
\phi_{tt}
+g\phi_z
+2\bar{\nu}\phi_{zzt}
+\left(|\nabla\phi|^2\right)_t
+\frac12\nabla\phi\cdot\nabla\!\left(|\nabla\phi|^2\right)
+2\bar{\nu}\nabla\phi\cdot\nabla\phi_{zz}\\
=
2\bar{\nu}
\left[
\phi_{txx}
+2\bar{\nu}\phi_{zzxx}
+\frac12\!\left(|\nabla\phi|^2\right)_{xx}
\right],
\qquad
z=\eta(x,t).
\label{2.8}
\end{split}
\ee

Equations \rf{2.4}, \rf{2.6}, \rf{2.7} and \rf{2.8}, form the starting point for the perturbation analysis.

For later   multiple-scales calculation, we expand equation \rf{2.8} 
to obtain 
\be
\begin{split}
0={}&
\phi_{tt}
+g\phi_z
+4\bar{\nu}\phi_{zzt}
+2\phi_x\phi_{xt}
+2\phi_z\phi_{zt}
+\phi_x^{2}\phi_{xx}
+2\phi_x\phi_z\phi_{xz}
+\phi_z^{2}\phi_{zz}\\
&\quad
+2\bar{\nu}
\left(
\phi_x\phi_{xzz}
+\phi_z\phi_{zzz}
\right)
\\
&\quad
-4\bar{\nu}^{\,2}
\left(
\phi_{xxzz}
\right)
-2\bar{\nu}
\left(
\phi_{xx}^2
+\phi_x \phi_{xxx}
+\phi_{xz}^2
+\phi_z \phi_{xxz}
\right),
\qquad
z=\eta(x,t).
\label{2.9}
\end{split}
\ee

The terms in equation \rf{2.9} naturally separate into linear, quadratic, cubic, and nonlinear viscous contributions, a decomposition which will be used in Section 3.

\section{Multiple-Scales Formulation}

The rapid oscillations of the carrier wave are represented by the phase
\begin{equation}
\theta=\omega t-kx,
\label{3.1}
\end{equation}
where $\omega^2=gk$ is the deep-water dispersion relation.
The envelope evolves on the slow  variables
\begin{equation}
X=\epsilon x,
\qquad
Z=\epsilon z,
\qquad
T=\epsilon t.
\label{3.2}
\end{equation}
Throughout the asymptotic analysis, we adopt the weak-viscosity scaling
\be
\bar{\nu}
=
\epsilon^{2}\nu,\label{3.3}
\ee
where \(\epsilon\ll1\) is the characteristic wave steepness and \(\nu=\mathcal O(1)\) with respect to \(\epsilon\). This scaling follows that of Carter and Govan \cite{CG2016} which  places the leading  viscous damping at nonlinear Schr\"odinger order.
The present calculation is  organized as a temporal-evolution equation rather than the spatial-evolution form used in their final nondimensionalization.

The analysis is carried out in the dimensional slow variables of the DDZ/Carter-Govan hierarchy. The nondimensional mean-flow-damped Dysthe equation introduced in the Introduction  serves only to identify the target complex mean-flow correction; the comparison concerns its asymptotic order and operator structure rather than a
term-by-term correspondence between the two normalizations.

\subsection{Harmonic Expansions}

The free-surface elevation and the velocity potential are expanded as
\begin{align}
\eta
&=
\epsilon Be^{i\theta}
+\epsilon^2B_2e^{2i\theta}
+\epsilon^3
\left(
\bar{\eta}
+B_3e^{3i\theta}
\right)
+\epsilon^4B_4e^{4i\theta}
+\cdots
+\mathrm{c.c.},
\label{3.4}
\\
\phi
&=
\epsilon Ae^{kz+i\theta}
+\epsilon^2
\left(
\bar{\phi}
+A_2e^{2(kz+i\theta)}
\right)
+\epsilon^3A_3e^{3(kz+i\theta)}
+\epsilon^4A_4e^{4(kz+i\theta)}
+\cdots
+\mathrm{c.c.}.
\label{3.5}
\end{align}
Here \(A(X,Z,T)\) and \(B(X,T)\) are the first-harmonic amplitudes of the velocity potential and free-surface elevation, respectively. The zero-harmonic quantities  \(\bar{\phi}(X,Z,T)\) and \(\bar{\eta}(X,T)\) represent the induced mean flow and mean free-surface displacement, while \(A_n\) and \(B_n\) denote the higher harmonic amplitudes.

The factors \(e^{nkz}\) provide the leading exponential decay of the oscillatory harmonics with depth. Their  slow \(Z\)-dependence  accounts for  modulation corrections required by Laplace's equation. The mean-flow potential \(\bar{\phi}\) has no rapidly varying vertical exponential factor and instead satisfies a separate slow-scale Laplace problem in the lower half-plane.

Substitution of the expansions into Laplace's equation and the Taylor-expanded free-surface boundary conditions produces a hierarchy of boundary-value problems at each order
\begin{equation}
\mathcal O(\epsilon^m e^{in\theta}).\label{3.6}
\end{equation}

At each perturbation order, the nonresonant higher harmonics are determined from boundary-value problems, the zero harmonic determines the induced mean flow, and the first harmonic satisfies a solvability condition that  yields the corresponding envelope evolution equation.

\subsection{Taylor Expansion of the Free-Surface Boundary Conditions}

The free-surface boundary conditions are imposed on the unknown surface
\(
z=\eta(x,t)
\)
and then Taylor expanded about the fixed boundary \(z=0\), according to
\begin{equation}
F(x,\eta,t)
=
F(x,0,t)
+\eta F_z(x,0,t)
+\frac12\eta^2F_{zz}(x,0,t)
+\cdots,
\label{3.7}
\end{equation}
where all derivatives are evaluated at \(z=0\).

\subsubsection{Taylor Expansion of the Kinematic Boundary Condition}
Applying \rf{3.7} to the kinematic boundary condition \rf{2.6} gives:
\be
\begin{split}
K(x,\eta,t)
&=
K + \eta K_z + \frac12\eta^2 K_{zz} + \cdots
\\
&=
\eta_t - \left(\phi_z + \eta \phi_{zz} + \half \eta^2 \phi_{zzz}\right)
+ \eta_x \left(  \phi_x + \eta \phi_{xz} + \half \eta^2 \phi_{xzz} \right)
- 2\bar{\nu} \eta_{xx} = 0.\label{3.8}
\end{split}
\ee

\subsubsection{Taylor Expansion of the Derivative Bernoulli Equation}
Likewise, if  ${\cal B}$ denotes the derivative Bernoulli equation introduced in Section~2, then 
\begin{equation}
{\cal B}(x,\eta,t)
=
{\cal B}
+\eta {\cal B}_z
+\frac12\eta^2{\cal B}_{zz}
+\cdots
=0,
\label{3.9}
\end{equation}
where, again, all quantities are evaluated at \(z=0\).

Using Laplace's equation and restricting attention to two-dimensional flow,
we decompose the derivative Bernoulli operator into its linear, quadratic, cubic, and nonlinear viscous blocks,
\begin{equation}
{\cal B}
=
L
+
Q_x
+
Q_z
+
C
+
V,
\label{3.10}
\end{equation}
where
\bea
L
&=&
\phi_{tt}
+g\phi_z
+4\bar{\nu}\phi_{zzt}
-4\bar{\nu}^{\,2}\phi_{xxzz},
\nonumber\\
Q_x
&=&
2\phi_x\phi_{xt},
\qquad
Q_z
=
2\phi_z\phi_{zt},
\nonumber\\
C
&=&
\phi_x^{2}\phi_{xx}
+2\phi_x\phi_z\phi_{xz}
+\phi_z^{2}\phi_{zz},
\nonumber\\
V
&=&
2\bar{\nu}
\left(
\phi_x\phi_{xzz}
+\phi_z\phi_{zzz}
\right)
-2\bar{\nu}
\left(
\phi_{xx}^{2}
+\phi_x\phi_{xxx}
+\phi_{xz}^2
+\phi_z\phi_{xxz}
\right).
\nonumber
\eea

The Taylor expansion \rf{3.7} is applied separately to each block $L$, $Q_x$, $Q_z$, $C$, and $V$, retaining  only the orders required in the particular
calculation under consideration. This approach avoids unnecessary algebra while ensuring that all contributions required at a given asymptotic order are retained.

\section{Weakly Viscous Perturbation Analysis Through Dysthe Order}

Through fourth order, the perturbation hierarchy recovers the weakly viscous Dysthe equation derived by Carter and Govan. The calculation also determines additional harmonic corrections that do not enter the fourth-order solvability condition but are required in the fifth-order analysis. These corrections are collected in Section 4.3.

As the perturbation hierarchy proceeds, each  harmonic amplitudes acquires
higher-order corrections. We write
\be
\begin{split}
A_n
&=
A_n^{(0)}
+\epsilon A_n^{(1)}
+\epsilon^2A_n^{(2)}
+\cdots,\\
B_n
&=
B_n^{(0)}
+\epsilon B_n^{(1)}
+\epsilon^2B_n^{(2)}
+\cdots,
\label{4.1}
\end{split}
\ee
where  \(A_n^{(0)}\) and \(B_n^{(0)}\) are  the leading \(n\)-th harmonic amplitudes, while \(A_n^{(j)}\) and \(B_n^{(j)}\) denote their \(j\)-th corrections. For the fundamental mode,
\begin{equation}
A_1^{(0)}=A,
\qquad
B_1^{(0)}=B.
\label{4.2}
\end{equation}

Similarly, the induced mean-flow potential is expanded as
\begin{equation}
\bar{\phi}
=
\bar{\phi}^{(0)}
+\epsilon\bar{\phi}^{(1)}
+\epsilon^2\bar{\phi}^{(2)}
+\cdots.
\label{4.3}
\end{equation}

\subsection{Lower-Order Asymptotics}
The leading orders determine the linear carrier-wave structure, envelope transport, the first nonlinear Stokes corrections, and the leading induced mean flow.
The leading-order first harmonic problem gives the deep-water dispersion relation,
\begin{equation}
  \omega^{2} = gk. \label{4.4}
\end{equation}
together with  the relation between the leading free-surface and potential amplitudes,
\begin{equation}
  B = -\frac{ik}{\omega}A. \label{4.5}
\end{equation}

At the next order, Laplace's equation gives $A_Z = i A_X$, while the resonant
first-harmonic problem  gives the transport equation
\begin{equation}
  A_T+ \frac{\omega}{2k}A_X =0.
  \label{4.6}
\end{equation}
Thus, the envelope propagates at the deep-water group velocity
$\displaystyle
c_g=\frac{\omega}{2k}.
$

The leading nonresonant second- and third-harmonic problems give
\begin{equation}
A_2^{(0)} =0,\qquad
B_2^{(0)} = kB^2,
\label{4.7}
\end{equation}
and 
\begin{equation}
A_3^{(0)} =0, \qquad
B_3^{(0)} = \frac{3}{2} k^2B^3  =  \frac{3i k^5}{2 \omega^3}A^3.
\label{4.8}
\end{equation}

 The zero harmonic at this order determines the leading induced mean flow.
 At the free surface,
\begin{equation}
\bar{\phi}_Z^{(0)}
=
\frac{2k^2}{\omega}
\left(|A|^2\right)_X,
\qquad
Z=0,
\label{4.9}
\end{equation}
while 
\begin{equation}
\bar{\phi}_{XX}^{(0)}+\bar{\phi}_{ZZ}^{(0)}
 =0,
\qquad
Z<0,
\label{4.10}
\end{equation}
with
$
\nabla\bar{\phi}^{(0)}
\longrightarrow0$ as
$
Z\longrightarrow-\infty$. Although  generated at third order, the mean-flow
first   feeds back into the envelope equation at fourth order.

The resonant  first harmonic solvability condition at \(\mathcal O(\epsilon^3e^{i\theta})\)
supplies the first nonlinear modulation correction,
\begin{equation}
\left(\frac{\omega}{2k}\right)^2A_{XX}
+4k^4|A|^2A
+4i\nu k^2\omega A
=0.
\label{4.11}
\end{equation}

Combining this  with the transport equation \rf{4.6} gives 
the  weakly viscous  nonlinear Schr\"odinger equation
\begin{equation}
2i\omega
\left(
A_T+\frac{\omega}{2k}A_X
\right)
+\epsilon
\left[
\left(\frac{\omega}{2k}\right)^2A_{XX}
+4k^4|A|^2A
+4i\nu k^2\omega A
\right]
=0.
\label{4.12}
\end{equation}
The first two terms inside the brackets are the standard dispersive and cubic nonlinear NLS contributions, while the final term is the leading viscous damping term.

\subsection {The Weakly Viscous Dysthe Equation}

At fourth order, the leading induced mean flow first couples back into the resonant first harmonic equation, producing the characteristic nonlocal interaction term  of Dysthe
theory. 
Solving the harmonic  mean-flow problem gives
\begin{equation}
  \bar{\phi}_X^{(0)} = -\frac{2k^2}{\omega} \mathcal{H} \!\left[ \left(|A|^2\right)_X \right], \qquad Z=0, \label{4.13}
\end{equation}
where \(\mathcal{H}\) denotes the Hilbert transform \cite{A2011}
This relation gives the nonlocal horizontal mean flow velocity at the free surface.

The resonant  first harmonic solvability condition at \(\mathcal O(\epsilon^4e^{i\theta})\)  is obtained
by eliminating the first harmonic corrections and substituting the lower-order relations derived in Section~4.1. Combining this result with the weakly viscous NLS equation \rf{4.12} yields
\be
\begin{split}
& 2i\omega
\left(
A_T + \frac{\omega}{2k}A_X
\right)
+
\epsilon
\left[
\left(\frac{\omega}{2k}\right)^2A_{XX}
+4k^4|A|^2A
+4i\nu k^2\omega A
\right]\\
&+
\epsilon^2
\left[
-\frac{8ik^3}{\omega^2}A_{XXX}
+16ik^3|A|^2A_X
-2ik^3A^2A_X^{*}
+2k\omega A\bar{\phi}^{(0)}_X
-8k\omega\nu A_X
\right]
=0.
\label{4.14}
\end{split}
\ee

The $\mathcal O(\epsilon^2)$ terms are  the higher-order dispersive, nonlinear, mean-flow, and viscous corrections characteristic of the weakly viscous Dysthe equation. In particular, the mean-flow interaction appears as $2k\omega A  \bar{\phi}_X^{(0)}$, with a purely real coefficient. Equation \rf{4.14}  recovers the
Carter-Govan viscous Dysthe equation and provides the fourth-order base for the higher-order analysis.

\subsection{Additional Lower-Order Harmonic Relations}

The perturbation hierarchy also yields several  additional lower-order
harmonic corrections that  do not enter  the fourth-order envelope equation but
are required in 
the fifth-order analysis. We collect these corrections in the order in which they arise.

The nonresonant first-harmonic problem at $\mathcal O(\eps^2 e^{i \theta})$ gives
\begin{equation}
A_1^{(1)}=0,
\qquad
B_1^{(1)}
=
\frac{1}{2\omega}A_X.
\label{4.15}
\end{equation}

Likewise, the second-harmonic problem at   \(\mathcal O(\epsilon^3e^{2i\theta})\) gives
\begin{align}
A_2^{(1)}&=0,
\quad B_2^{(1)}
=
-\frac{2ik}{g}AA_X.
\label{4.16}
\end{align}

At  ${\cal O}(\epsilon^4 \e^{2i\theta})$, the second-harmonic problem yields
\begin{equation}
A_2^{(2)}
=
\frac{2\nu k^4}{\omega^2}A^2
-\frac{i}{8\omega}AA_{XX}
-\frac{6ik^6}{\omega^3}|A|^2A^2.
\label{4.17}
\end{equation}

The correction $A_2^{(2)}$
is not needed in  the fourth-order Dysthe solvability condition.
It is required at fifth order, however,  because the corrected second harmonic can interact with the conjugate fundamental mode to produce a resonant first-harmonic forcing.
The corresponding surface-elevation coefficient \(B_2^{(2)}\)
does not enter the fifth-order envelope equation, although it would be needed to reconstruct the free-surface to the same asymptotic order.

 The nonresonant problem at $\mathcal O(\eps^4 e^{4 i \theta})$  determines the leading fourth harmonic. The  lower-order harmonic relations required in the fifth-order calculation are summarized as follows:

\begin{equation}
\begin{split}
B
&=
-\frac{ik}{\omega}A,
\\
A_1^{(1)}
&=
0,
\qquad
B_1^{(1)}
=
\frac{1}{2\omega}A_X
\\
A_2^{(0)}
&=
A_2^{(1)}
=0,
\qquad
B_2^{(0)}
=
kB^2,
\qquad
B_2^{(1)}
=
-\frac{2ik}{g}AA_X,
\\
A_2^{(2)}
&=
\frac{2\nu k^4}{\omega^2}A^2
-\frac{i}{8\omega}AA_{XX}
-\frac{6ik^6}{\omega^3}|A|^2A^2,
\\
A_3^{(0)}
&=
0,
\qquad
B_3^{(0)}
=
\frac{3}{2}k^2B^3
\\
A_4^{(0)}
&=
0,
\qquad
B_4^{(0)}
=
\frac{8}{3} k^3 B^4.
\label{4.18}
\end{split}
\end{equation}

Equation \rf{4.18} completes the lower-order  hierarchy required for the fifth-order harmonic solvability condition.
Higher corrections to the third harmonic, and additional  fourth- and fifth-harmonic amplitudes, are not required for the fifth-order first-harmonic solvability condition by order and harmonic counting. They enter only if one reconstructs the full wave profile rather than the envelope equation.

\section{Fifth Order Extension}

The leading Dysthe mean-flow interaction enters at fourth order. Under the
weak-viscosity scaling $\bar\nu = \eps^2 \nu$, multiplication of this interaction by an explicit factor of viscosity would raise it to sixth order. This
order-counting argument excludes a direct viscous correction at fifth order, but it does not exclude an indirect contribution generated through viscosity-dependent lower-order corrections.

Viscosity already enters the perturbation hierarchy through the weakly viscous NLS equation and through harmonic corrections such as $A_2^{(2)}$. These
quantities participate in the nonlinear interactions contributing to the
fifth-order first-harmonic solvability condition and could, in principle, generate the same operator structure as the prescribed complex mean-flow correction. We therefore calculate the first correction to the induced mean flow and the complete fifth-order first-harmonic solvability condition.

\subsection{First Correction to the Induced Mean Flow}
The first correction to the induced mean-flow potential is determined from the zero-harmonic problem at $\mathcal O(\epsilon^4)$.
 For convenience, define 
\begin{equation}
J =A^{*}A_{XX}-AA_{XX}^{*} \label{5.1}
\end{equation}
The surface boundary condition for \(\bar{\phi}^{(1)}\) is then 
\begin{equation}
  \bar{\phi}_Z^{(1)} = \frac{k}{2\omega} \mathcal{H} \!\left[ |A|^2_{XX} \right] + \frac{5ik}{4\omega}J, \qquad Z=0. \label{5.2}
\end{equation}

Since \(\bar{\phi}^{(1)}\) is harmonic for \(Z<0\) and decays as
\(Z\rightarrow-\infty\), its surface  derivatives satisfy 
\begin{equation}
\bar{\phi}_X^{(1)}
=
-\mathcal{H}
\!\left[
\bar{\phi}_Z^{(1)}
\right],
\qquad
Z=0.\label{5.3}
\end{equation}

Applying this relation to the boundary data in \rf{5.2} gives
\begin{equation}
  \bar{\phi}_X^{(1)} = \frac{k}{2\omega}|A|^2_{XX} - \frac{5ik}{4\omega} \mathcal{H}[J], \qquad Z=0. \label{5.4}
\end{equation}

Since $J$ is purely imaginary, \(\mathcal{H}[J]\) is also purely imaginary,
and 
\(\bar{\phi}_X^{(1)}\) is therefore real-valued, as required for the horizontal induced mean-flow velocity at the free surface.

\subsection{Fifth-Order First-Harmonic Solvability Condition}

The fifth-order envelope equation follows from the  solvability
condition for the 
\(\mathcal O(\epsilon^5e^{i\theta})\) coefficient in the derivative Bernoulli equation.
Using  the decomposition introduced in
Section~3,
\begin{equation}
\mathcal B = L+Q_x+Q_z+C+V.
\label{5.5}
\end{equation}
we isolate the first-harmonic contribution of each block at fifth order and substitute the lower-order harmonic relations from Section 4.3. In particular, retaining the viscosity-dependent correction $A_2^{(2)}$ includes the indirect viscous contributions transmitted through the corrected second harmonic. Mean-flow derivatives are kept explicit until the complete fifth-order forcing has been assembled.

The  envelope equation through fifth order may be written as
\begin{equation}
2i\omega
\left(
A_T+\frac{\omega}{2k}A_X
\right)
+\epsilon F_3
+\epsilon^2F_4
+\epsilon^3F_5
=0,
\label{5.6}
\end{equation}
where \(F_3\) and \(F_4\) are, respectively,  the weakly viscous NLS 
and fourth-order Dysthe contributions  obtained in Section~4.
The new fifth-order contribution is $F_5$.

Before the mean-flow derivatives are eliminated using the relations
from  Section 5.1,
 the fifth-order forcing is
 \be
\begin{split}
F_5={}&
\frac{13}{8}k^2A^2A_{XX}^{*}
-\frac{21}{4}k^2|A|^2A_{XX}
+14k^2AA_XA_X^{*}
-6k^2A^{*}A_X^2
\\
&\quad
+\frac{30k^8}{\omega^2}|A|^4A
+\frac{2i\nu k^6}{\omega}|A|^2A
+4\nu^2k^4A
-8i\nu\omega A_{XX}
\\
&\quad
+2k\omega A\bar{\phi}_X^{(1)}
+2ik\omega A\bar{\phi}_Z^{(1)}
-2ikA\bar{\phi}_{XT}^{(0)}
+2kA\bar{\phi}_{ZT}^{(0)}
\\
&\quad
-i\omega A\bar{\phi}_{ZZ}^{(0)}
+3i\omega A_X\bar{\phi}_X^{(0)}
-3\omega A_X\bar{\phi}_Z^{(0)}.\label{5.7}
\end{split}
\ee

The leading transport equation implies
\begin{equation} |A|^2_T = -\frac{\omega}{2k}|A|^2_X.
  \label{5.8}
\end{equation}
Differentiating  the leading mean-flow relations  with respect to $T$ therefore gives 
\be
\bar{\phi}_{XT}^{(0)} = k\,\mathcal{H}[|A|^2_{XX}], \qquad \bar{\phi}_{ZT}^{(0)} = -k|A|^2_{XX}.
\label{5.9}
\ee

Laplace's equation also gives
\begin{equation}
  \bar{\phi}_{ZZ}^{(0)} = \frac{2k^2}{\omega} \mathcal{H}[|A|^2_{XX}].\label{5.10}
\end{equation}
Substituting these identities, together with the first-order mean-flow relations \rf{5.2}--\rf{5.4}, into equation \rf{5.7} yields the  reduced fifth-order contribution,
\be
\begin{split}
F_5={}&
\frac{25}{8}k^2A^2A_{XX}^{*}
-\frac{35}{4}k^2|A|^2A_{XX}
+6k^2AA_XA_X^{*}
-12k^2A^{*}A_X^2
\\
&\quad
+\frac{30k^8}{\omega^2}|A|^4A
+\frac{2i\nu k^6}{\omega}|A|^2A
+4\nu^2k^4A
-8i\nu\omega A_{XX}
\\
&\quad
-3ik^2A\mathcal{H}[|A|^2_{XX}]
-\frac{5}{2}ik^2A\mathcal{H}[J]
-6ik^2A_X\mathcal{H}[|A|^2_X].\label{5.11}
\end{split}
\ee

Equations \rf{5.6} and \rf{5.11}, together with the previously determined expressions for
$F_3$ and $F_4$,  constitute the envelope equation through fifth order.

The terms in $F_5$ separate naturally according to their structure.
The first four terms in \rf{5.11} are higher-order nonlinear dispersive contributions involving second derivatives of the envelope and products of envelope gradients. They are followed by a quintic self-interaction, three  local viscous corrections, and  three nonlocal contributions
associated with the induced mean flow.

The fifth-order nonlocal terms differ structurally from the leading Dysthe mean-flow interaction. At fourth order, the induced mean flow enters through
$
A\bar{\phi}_X^{(0)},
$
or, after solving the mean-flow problem, through
$
A\mathcal{H}[|A|^2_X].
$

At  fifth order  the corresponding nonlocal terms are 
\begin{equation}
A\mathcal{H}[|A|^2_{XX}],
\qquad
A_X\mathcal{H}[|A|^2_X],
\qquad
A\mathcal{H}[J].
\label{5.12}
\end{equation}

The first term represents a higher-order response to  the
spatial curvature of the wave intensity, while 
he second  couples the local envelope slope to the leading nonlocal mean-flow response.
For the third term,  writing $ A = Re^{i\vartheta}$, gives
\begin{equation}
  J= A^*A_{XX} -A A^*_{XX} = 2i\left(R^2\vartheta_X\right)_X,\label{5.13}
  \end{equation}
so that $A\mathcal{H}[J]$ represents a nonlocal coupling to  gradients of the local phase current. 
Thus, the fifth-order mean-flow feedback introduces new derivative-dependent
nonlocal structures rather than changing only  the coefficient of the leading Dysthe operator.

The explicit $\nu$- and $\nu^2$-dependent terms in \rf{5.11} show that
viscosity dependent lower-order corrections are fully represented in the
the fifth-order solvability calculation.
However,  none of the nonlocal terms in \rf{5.12} carries an explicit factor
of $\nu$. Consequently, the fifth-order contribution \rf{5.11} contains no term proportional to the prescribed form 
\begin{equation} i\nu A\bar{\phi}_X^{(0)}, \label{5.14}
\end{equation} or, equivalently, \begin{equation} i\nu A\mathcal{H}[|A|^2_X]. \label{5.15} \end{equation}

The absence of this contribution follows from  the complete fifth-order calculation, including the viscosity-dependent harmonic corrections and lower-order evolution equations, rather than from order counting alone. Within the DDZ reduction, the fifth-order terms modify the functional form of the induced mean-flow feedback but do not generate the phenomenological modification
\begin{equation} A\mathcal{H}[|A|^2_X] \;\longmapsto\; A(1+i\beta)\mathcal{H}[|A|^2_X]. \label{5.16} \end{equation}

\section{Sixth-Order Direct Viscous Contribution to  the Mean-Flow Interaction}

By order counting, under the weak-viscosity scaling $ \bar{\nu}=\epsilon^2\nu,$ a direct viscous
correction to the fourth-order Dysthe mean-flow interaction first enters at
sixth order.
We therefore  isolate the corresponding carrier-mean contribution from the nonlinear viscous block without deriving the complete sixth-order envelope equation. This restricted calculation determines the asymptotic order and operator structure of the contribution, but not its net coefficient in the complete sixth-order first-harmonic solvability condition.

\subsection{Carrier-Mean Contribution from the Nonlinear Viscous Block}

To  isolate the interaction between the leading carrier wave and the leading induced mean flow,  write the relevant part of the velocity potential as
\begin{equation}
  \phi = \phi^{c} + \phi^{m} +\cdots, \label{6.1}
\end{equation}
where
\begin{align} \phi^{c} &= \epsilon A(X,Z,T)e^{kz+i\theta} +\mathrm{c.c.}, & \phi^{m} &= \epsilon^2\bar{\phi}^{(0)}(X,Z,T). \label{6.2} \end{align}
The omitted terms consist of higher harmonics and higher-order corrections that do not enter the direct carrier--mean contribution considered here.

Since the nonlinear viscous block is quadratic in \(\phi\), substitution of \rf{6.1} separates its terms  into carrier--carrier, mean--mean, and carrier--mean
contributions
Here, a direct carrier--mean contribution is a cross term that is linear in both \(\phi^{c}\) and \(\phi^{m}\) and carries an explicit factor of \(\bar{\nu}\).
Since \(\phi^{m}\) is a zero-harmonic component and \(\phi^{c}\) contains the first harmonic, their product can contribute directly to the first-harmonic solvability condition.
This mechanism is distinct from the indirect fifth-order mechanism examined in Section~5, where viscosity enters through lower-order harmonic corrections and evolution equations.

The mean flow depends only on the slow variables \(X\) and \(Z\), so  one
physical spatial derivative of $\phi_x^{m}$ introduces an additional factor of \(\epsilon\). Thus, $\phi_x^{m}$ and $\phi_z^{m}$ are both $\mathcal O(\epsilon^3)$.

By contrast, third-order spatial derivatives of  \(\phi^{c}\) are dominated by
derivatives of the rapidly varying factor $e^{kz+i\theta}$ and
remain $\mathcal O(\epsilon)$.
A product containing one derivative of \(\phi^{m}\) and one third-order  derivative of \(\phi^{c}\) is therefore \(\mathcal O(\epsilon^4)\).
Multiplication by $\bar\nu$ places the resulting direct carrier--mean interaction  at $\mathcal O(\epsilon^6)$.

The nonlinear viscous block of the derivative Bernoulli equation, where $ \bar{\nu}=\epsilon^2\nu $, is
\be
\begin{split}
  V={}& 2\bar{\nu} \left( \phi_x\phi_{xzz} + \phi_z\phi_{zzz} \right)
  \\
  &\quad - 2\bar{\nu} \left( \phi_{xx}^{\,2} + \phi_x\phi_{xxx} + \phi_{xz}^{\,2} + \phi_z\phi_{xxz} \right). \label{6.3}
\end{split}
\ee
Substitution of  \rf{6.1}, into each quadratic term in \(V\) produces carrier-carrier, mean-mean, and  carrier-mean cross terms. Only some of the carrier-mean terms enter at sixth order.
For example,
\begin{equation} \phi_{xx}^{\,2} = \left(\phi_{xx}^{c}\right)^2 + 2\phi_{xx}^{c}\phi_{xx}^{m} + \left(\phi_{xx}^{m}\right)^2 +\cdots.\label{6.4}
\end{equation}
Although  $2\phi_{xx}^{c}\phi_{xx}^{m}$ is a carrier--mean term, it does not contribute at sixth order because $\phi_{xx}^{m} = \mathcal O(\epsilon^4)$ and $\phi_{xx}^{c} = \mathcal O(\epsilon)$.  After multiplication by $\bar{\nu}$, this term is \(\mathcal O(\epsilon^7)\).

The same argument excludes carrier--mean terms in which the mean-flow factor carries two or more physical spatial derivatives.
The sixth order contribution therefore consists only of terms  in which the mean flow carries one spatial derivative and the carrier carries a third-order
derivative.

The required derivatives of the leading carrier are
\be
\begin{split}
  \phi_{xzz}^{c} &= -ik^3\epsilon A e^{kz+i\theta} + \mathcal O(\epsilon^2), \qquad \phi_{zzz}^{c} = k^3\epsilon A e^{kz+i\theta} + \mathcal O(\epsilon^2), \\
  \phi_{xxx}^{c} &= ik^3\epsilon A e^{kz+i\theta} + \mathcal O(\epsilon^2), \qquad \phi_{xxz}^{c} = -k^3\epsilon A e^{kz+i\theta} + \mathcal O(\epsilon^2).\label{6.5}
\end{split}
\ee

Let $V_{cm}$ denote the  sixth-order part of  V that is linear in both
$\phi^{m}$ and $\phi^{c}$. Retaining the relevant cross terms gives
\be
\begin{split} V_{\mathrm{cm}} ={}& 2\bar{\nu} \left( \phi_x^{m}\phi_{xzz}^{c} + \phi_z^{m}\phi_{zzz}^{c} \right)
  \\
  &\quad - 2\bar{\nu} \left( \phi_x^{m}\phi_{xxx}^{c} + \phi_z^{m}\phi_{xxz}^{c} \right). \label{6.6}
\end{split}
\ee

Substituting the leading carrier derivatives from \rf{6.5} and the
mean flow derivatives implied   \rf{6.2},
into equation \rf{6.6}, we find that the two bracketed groups of terms contribute equally to the coefficient of $\epsilon^6e^{i\theta}$. This equality holds only at the order retained here. Evaluating at \(z=0\) gives
\begin{equation} V_{\mathrm{cm}} = 4\epsilon^6\nu k^3 A e^{i\theta} \left( \bar{\phi}_Z^{(0)} -i\bar{\phi}_X^{(0)} \right) +\mathrm{c.c.} + \mathcal O(\epsilon^7), \qquad Z=0.
  \label{6.7}
\end{equation}

The leading induced mean flow is harmonic in the lower half-plane and decays as \(Z\rightarrow-\infty\).
Substituting its surface derivatives from equations \rf{4.9} and \rf{4.13} into equation \rf{6.7} yields
\begin{equation}
  V_{\mathrm{cm}} = \frac{8\epsilon^6\nu k^5}{\omega} A e^{i\theta} \left\{ \left(|A|^2\right)_X +i\mathcal{H} \!\left[ \left(|A|^2\right)_X \right] \right\} +\mathrm{c.c.} + \mathcal O(\epsilon^7). \label{6.8}
\end{equation}

The component proportional to $\left(|A|^2\right)_X$ in \rf{6.8} is local.
The nonlocal component of the nonlinear viscous block contains is
\begin{equation}
  \frac{8i\nu k^5}{\omega} A\mathcal{H} \!\left[ \left(|A|^2\right)_X \right].
  \label{6.9}
\end{equation}
This term has the operator structure of the presribed purely  imaginary  correction to the leading Dysthe mean-flow interaction. It differs fundamentally from the fifth-order nonlocal terms obtained in Section~5, which are derivative-dependent and contain no explicit factor of \(\nu\).  Equation \rf{6.9}, by contrast,  arises directly from the \(\bar{\nu}\)-weighted interaction between the leading carrier and the leading induced mean flow.
 
The complete sixth-order first-harmonic solvability condition has not been assembled, and other sixth-order contributions may possess the same operator structure and modify its coefficient. The present calculation therefore establishes
the asymptotic order and operator structure of the direct contribution generated by the nonlienar viscous block, but not its net  coefficient in the  complete sixth-order envelope equation.

\subsection{Interpretation and Scope of the DDZ Weak Viscosity Reduction}

The fifth- and sixth-order results should be interpreted within the DDZ weak-viscosity reduction summarized in Section 2.2. In this formulation, the leading
effect of near-surface rotational motion is represented through viscous corrections to the free-surface boundary conditions, while the velocity potential remains harmonic in the fluid interior. The preceding results
therefore apply only to the nonlinear interactions retained within this reduced system.

This restriction  is important when comparing the DDZ  model with formulations that retain additional rotational effects.
Eeltink et al. compared several descriptions of viscous free-surface flow and
examined the roles of rotational surface velocity and rotational pressure in potential-flow closures \cite{EABK2020}.
They found that  selected nonlinear vortical contributions to the free-surface boundary conditions cancel through $\mathcal O(\epsilon^4)$, leaving the viscous Dysthe equation unchanged. Their analysis does not determine whether this cancellation persists at fifth or sixth order. Extending such a formulation beyond Dysthe order would therefore provide a natural test of whether additional rotational effects modify either the fifth-order mean-flow interaction or the direct sixth-order contribution
identified here.

The DDZ reduction does not  resolve dissipative processes such as intermittent wave breaking and the associated turbulence, either of which could alter the induced mean-flow response. A separate possibility is that momentum transfer between the carrier wave and the induced current occurs over a finite adjustment time. In that case, the mean flow does not respond instantaneously to changes in the wave envelope. Such a delay would generate  a
phase lagged component of the mean-flow response. The physical origin of such a delay is not derived here. The reduced relaxation model introduced in Section~7 is intended only to examine its mathematical consequences.

\section{A Reduced Phase-Lag Model for Mean-Flow Feedback}

Earlier coupled-mode studies suggest that an induced low-frequency component may possess dissipative dynamics distinct from those of the carrier wave. In particular, Fabrikant retained a dynamically evolving low-frequency mode before eliminating it through a quasistatic approximation \cite{F1984}.
Motivated by this possibility, we introduce an independent reduced model in
which the induced mean flow adjusts to the instantaneous Dysthe prediction
over a finite time. The model is not derived from the DDZ hierarchy or from
Fabrikant's theory, and it is not intended to replace a higher-order
calculation retaining vortical dynamics. Its purpose is only to show how
finite-time adjustment generates a phase lag in the  mean-flow response and, in an appropriate low-frequency limit, a correction of the form \[ A(1+i\beta)M_0. \]

Define 
\begin{equation}
  M_0(X,T) = \bar{\phi}_X^{(0)}(X,0,T) = -\frac{2k^2}{\omega} \mathcal{H} \!\left[ \left(|A(X,T)|^2\right)_X \right]. \label{7.1}
\end{equation}
Here, \(A(X,T)\) is  the complex carrier-wave envelope,  $X=\epsilon x $
and $T=\epsilon t$  are the slow variables and
\(\mathcal{H}\) denotes the Hilbert transform with respect to \(X\).
Thus \(M_0(X,T)\) is the horizontal  mean-flow velocity at the free surface predicted by the instantaneous Dysthe mean-flow problem.

Let \(M(X,T)\) denote an effective horizontal mean-flow velocity that may include corrections to the instantaneous Dysthe response arising from unresolved rotational, turbulent, or pre-breaking dynamics.
We assume  that, at each fixed $X$,  \(M\) relaxes toward  \(M_0(X,T)\)  according to 
\begin{equation} \tau M_T+M=M_0. \label{7.2}
\end{equation}
The parameter \(\tau>0\) is a specified adjustment time measured on the slow time scale \(T\).
In the limit $ \tau\rightarrow 0$, equation \rf{7.2} reduces to $ M=M_0$,
recovering the instantaneous Dysthe mean-flow response.

First-order relaxation laws of the general form \[ \tau q_t+q=q_{\mathrm{eq}} \] are widely used in continuum mechanics and fluid modeling  when a flux, stress, or other response variable approaches an equilibrium value over a finite time rather than instantaneously. Classical examples include Maxwell stress relaxation
and  Maxwell--Cattaneo-type models, which employ analogous finite-time adjustment laws for transport fluxes \cite{HPE2021}.
Related relaxation ideas also arise in  models coupling  waves to a  mean-flow.
In particular, M\"uller incorporated relaxation of an internal-wave field coupled to a larger-scale oceanic mean flow \cite{M1976}. These earlier studies provide a modeling rationale for the mathematical form of equation \rf{7.2}, but they do not establish its specific application to the induced mean flow considered here.

To determine the temporal response  of equation \rf{7.2},
consider a Fourier
component of the instantaneous mean flow on the slow time scale,
\begin{equation}
  M_0(X,T) = \widehat{M}_0(X;\Omega)e^{-i\Omega T}.
  \label{7.3}
\end{equation}
Here \(\widehat{M}_0(X;\Omega)\) is the spatial amplitude associated with the
slow modulation frequency \(\Omega\). Because \rf{7.2} is linear and contains no derivatives with respect to $X$, the response may be analyzed independently at each fixed $X$. A general forcing can then be constructed by superposition
whenever the corresponding  Fourier  representation is valid.


We seek a response at the same frequency, 
\begin{equation} M(X,T) = \widehat{M}(X;\Omega)e^{-i\Omega T}. \label{7.4}
\end{equation}
Then
\begin{equation} M_T = -i\Omega\widehat{M}(X;\Omega)e^{-i\Omega T}.\label{7.5}
\end{equation}
Substitution of equations \rf{7.3} - \rf{7.5} into equation \rf{7.2},
followed by cancellation of the common factor \(e^{-i\Omega T}\), gives
\begin{equation}
  (1-i\Omega\tau)\widehat{M} = \widehat{M}_0.\label{7.6}
\end{equation}

Hence, \begin{equation}
  \frac{\widehat{M}(X;\Omega)}{\widehat{M}_0(X;\Omega)} = \frac{1}{1-i\Omega\tau} = \frac{1+i\Omega\tau}{1+\Omega^2\tau^2}.
  \label{7.7}
\end{equation}

The transfer function of the relaxation model is then 
\begin{equation}
  G(\Omega) = \frac{\widehat{M}(X;\Omega)} {\widehat{M}_0(X;\Omega)} = \frac{1}{1-i\Omega\tau}.
\end{equation}
It specifies the amplitude and phase of the mean-flow response relative to the instantaneous forcing at the modulation frequency \(\Omega\).
In polar form,
\begin{equation}
  G(\Omega) = \frac{1}{\sqrt{1+\Omega^2\tau^2}} e^{i\delta(\Omega)}, \label{7.8}
\end{equation}
where
\begin{equation} \delta(\Omega) = \tan^{-1}(\Omega\tau). \label{7.9}
\end{equation}
Thus, the amplitude of each temporal Fourier component is reduced by the factor \[ |G(\Omega)| = \left(1+\Omega^2\tau^2\right)^{-1/2}, \] while its phase is shifted by \(\delta(\Omega)\).

Define the equivalent time delay $\Delta T(\Omega)$, for $\Omega \neq 0$ by 
$e^{i\delta(\Omega)}e^{-i\Omega T}
=
e^{-i\Omega\left[T-\Delta T(\Omega)\right]}$. It follows that 
\begin{equation}
  \Delta T(\Omega) = \frac{\delta(\Omega)}{\Omega} =
  \frac{\tan^{-1}(\Omega\tau)}{\Omega}.\label{7.10}
\end{equation}
The corresponding real-valued response is therefore delayed relative to the forcing by
$ \Delta T(\Omega)$. In the low-frequency limit, $ \Delta T(\Omega) \rightarrow \tau$.

In the low-frequency regime $|\Omega\tau|\ll 1$,
the transfer function has the expansion
\begin{equation}
    G(\Omega)  = 1+i\Omega\tau +O\!\left((\Omega\tau)^2\right). \label{7.11}
\end{equation}

Thus, the mean-flow feedback for a single Fourier component becomes
\begin{equation}
A\widehat{M}
\approx
A(1+i\Omega\tau)\widehat{M}_0.
\label{7.12}
\end{equation}
Since
\[
M_0
=
-\frac{2k^2}{\omega}
\mathcal{H}\!\left[(|A|^2)_X\right],
\]
the low-frequency response modifies the instantaneous Hilbert-transform
mean-flow interaction according to
\begin{equation}
A\mathcal{H}\!\left[(|A|^2)_X\right]
\longmapsto
A\left[1+i\beta_{\mathrm{eff}}(\Omega)\right]
\mathcal{H}\!\left[(|A|^2)_X\right],
\label{7.13}
\end{equation}
where
\begin{equation}
\beta_{\mathrm{eff}}(\Omega)=\Omega\tau.
\label{7.14}
\end{equation}

The relaxation model therefore does not generally produce  a constant value of \(\beta\).
Instead, the effective coefficient $ \beta_{\mathrm{eff}}(\Omega)$
varies across the temporal modulation spectrum. Approximating this
response by a constant \(\beta\) is justified only when  the relevant dynamics are concentrated within a sufficiently narrow frequency band centered on a representative modulation frequency \(\Omega_{*}\). Under this assumption,
\begin{equation}
  \beta \approx \Omega_{*}\tau. \label{7.15}
\end{equation}
The relaxation model does not determine the value of \(\Omega_{*}\).
Instead, it must  be associated with the dominant modulation scale of the wave field or specified through calibration.

The relation \(\beta_{\mathrm{eff}}(\Omega) = \Omega \tau\) also makes the asymptotic size of the phase-lag correction explicit. Since the instantaneous
Dysthe mean-flow feedback first enters at fourth order, 
imposing
\begin{equation}
  \Omega\tau=\mathcal O(\epsilon), \quad  \beta_{\mathrm{eff}}(\Omega) = \mathcal O(\epsilon)
  \label{7.16}
\end{equation}
places the correction $ i\Omega\tau\,A \widehat{M}_0 $ at fifth-order, one order beyond the leading Hilbert-transform mean-flow interaction.
Here, \(\Omega\) and \(\tau\) are measured on the slow temporal scale \(T\). This ordering is introduced solely to compare  the relaxation correction
with the  fifth order perturbation hierarchy. It is not derived from the
DDZ weak-viscosity reduction or from an independent physical estimate of the mean-flow adjustment time.

This model is deliberately minimal. It does not identify the physical process that determines the adjustment time $\tau$ or establish that vorticity, turbulence, or incipient wave breaking produces  the relaxation law \rf{7.2}.
Equation  \rf{7.2} should therefore be viewed as a reduced representation of unresolved delayed mean-flow dynamics rather than as a first-principles derivation. Establishing its physical basis and determining $\tau$ require a more complete theory that retains  the relevant nonlinear rotational or dissipative effects.

\section{Conclusions}
This  work extends the DDZ/Carter-Govan weak viscosity hierarchy beyond Dysthe order to determine whether it generates the prescribed imaginary  correction to
the nonlocal  mean-flow interaction.  
The complete fifth-order first-harmonic solvability condition produces
new derivative-dependent nonlocal terms but no viscosity-dependent term proportional to  $ i\nu A\bar{\phi}_X^{(0)}$. The proposed indirect fifth-order mechanism is therefore absent within the DDZ reduction.

At sixth order, the nonlinear viscous block contains a direct carrier--mean contribution proportional to
$ i\nu A\mathcal{H} \!\left[ \left(|A|^2\right)_X \right]$,
which has  the operator structure of the prescribed purely imaginary
correction to the leading Dysthe mean-flow interaction. Because the complete sixth-order solvability condition has not been assembled, the calculation determines the asymptotic order and operator structure of this contribution, but not its final coefficient in a complete sixth-order envelope equation.

These results apply specifically to the DDZ reduced potential-flow formulation and do not include the full nonlinear rotational dynamics of the free-surface
problem, turbulence, or wave breaking. Independently of the asymptotic
derivation,  the finite-adjustment-time model shows that  delayed mean-flow response produces the exact transfer function   $G(\Omega) = \frac{1}{1-i\Omega\tau}$,
whose  low-frequency expansion modifies the instantaneous Hilbert-transform
mean-flow interaction according to
\[
A\mathcal{H}\!\left[(|A|^2)_X\right]
\longmapsto
A\left[1+i\beta_{\mathrm{eff}}(\Omega)\right]
\mathcal{H}\!\left[(|A|^2)_X\right],
\qquad
\beta_{\mathrm{eff}}(\Omega)=\Omega\tau.
\]
Under the imposed ordering $\Omega\tau = \mathcal O(\epsilon)$, the imaginary phase-lag correction is $\mathcal O(\epsilon)$ relative to the leading Hilbert- transform mean-flow interaction and therefore enters at fifth order.
Determining whether such delayed dynamics arise from wave breaking, turbulence, rotational effects, or other physical processes omitted by the DDZ reduction, and determining the adjustment time
$\tau$, require a more complete model of the relevant dissipative dynamics.

\section*{Acknowledgements:} We gratefully acknowledge partial support for this research by the Simons Foundation through award \# 527565 (PI: C.M. Schober).

\bibliographystyle{plain}

\end{document}